\documentclass[a4paper,onecolumn,11pt,accepted=2021-06-18]{quantumarticle}
\pdfoutput=1
\usepackage[utf8]{inputenc}
\usepackage[english]{babel}
\usepackage[T1]{fontenc}
\usepackage{amsmath}
\usepackage{hyperref}

\usepackage{tikz}
\usepackage{lipsum}

\usepackage{hyperref}

\usepackage[normalem]{ulem}
\usepackage{cancel}
\newcommand{\strikered}[1]{\ifmmode\textcolor{red}{\cancel{#1}}\else\textcolor{red}{\sout{#1}}\fi}

\usepackage[left=1in,right=1in,top=1in,bottom=1in]{geometry}

\usepackage[noblocks]{authblk}
\usepackage{comment}
\usepackage{bm}
\usepackage{enumerate}
\usepackage{amsfonts}

\usepackage{amsmath}
\usepackage{graphicx}
 
\usepackage{mathtools}
\mathtoolsset{showonlyrefs}

\usepackage{dcolumn}

\usepackage{enumerate}
\usepackage{amssymb}
\usepackage{amsthm}
\usepackage{calc} 
\usepackage{accents}
\usepackage{dsfont}

\newcommand*{\cA}{\mathcal{A}} 
\newcommand*{\cB}{\mathcal{B}}
\newcommand*{\cC}{\mathcal{C}}

\newcommand*{\cH}{\mathcal{H}}

\newcommand*{\cJ}{\mathcal{J}}
\newcommand*{\cK}{\mathcal{K}}

\newcommand*{\cD}{\mathcal{D}}

\newcommand*{\cP}{\mathcal{P}}
\newcommand*{\cS}{\mathcal{S}}
\newcommand*{\cU}{\mathcal{U}}
\newcommand*{\cT}{\mathcal{T}}

\newcommand*{\tr}{\mathop{\mathrm{tr}}\nolimits}

\newcommand{\ran}{\rangle}
\newcommand{\lan}{\langle}

\makeatletter
\newtheorem*{rep@theorem}{\rep@title}
\newcommand{\newreptheorem}[2]{
\newenvironment{rep#1}[1]{
 \def\rep@title{#2 \ref{##1}}
 \begin{rep@theorem}}
 {\end{rep@theorem}}}
\makeatother

\newtheorem{theorem}{Theorem}
\newreptheorem{theorem}{Theorem}
\newtheorem{lemma}[theorem]{Lemma}
\newreptheorem{lemma}{Lemma}
\newtheorem{definition}[theorem]{Definition}
\newreptheorem{definition}{Definition}
\newtheorem{remark}[theorem]{Remark}
\newreptheorem{remark}{Remark}
\newtheorem{corollary}[theorem]{Corollary}
\newreptheorem{corollary}{Corollary}
\newtheorem{proposition}[theorem]{Proposition}
\newreptheorem{proposition}{Proposition}
\newtheorem{example}[theorem]{Example}
\newreptheorem{example}{Example}

\newcommand{\myacknowledgements}{\begin{center}{\bf Acknowledgements}\end{center}\par}

\usepackage{amsfonts}

\newcommand{\ket}[1]{|#1\rangle}
\newcommand*{\mylabel}[1]{\label{#1}}
\newcommand{\bra}[1]{\langle#1|}

\newcommand{\proj}[1]{|#1\rangle\!\langle#1|}

\newcommand{\assign}{:=}

\usepackage[symbol]{footmisc}

\numberwithin{equation}{section}

\usepackage[numbers,sort&compress]{natbib}

\begin{document}

\title{Programmability of covariant quantum channels}

\author{Martina Gschwendtner}
\affiliation{Munich Center for Quantum Science and Technology (MCQST), 80799 M\"unchen, Germany}
\affiliation{Zentrum Mathematik, Technical University of Munich, 85748 Garching, Germany}
\orcid{0000-0002-2155-9693}
\email{martina.gschwendtner@tum.de}
\author{Andreas Bluhm}
\email{bluhm@math.ku.dk}
\orcid{0000-0003-4796-7633}
\affiliation{QMATH, Department of Mathematical Sciences, \protect\\University of Copenhagen, 2100 Copenhagen, Denmark}
\author{Andreas Winter}
\email{andreas.winter@uab.cat}
\affiliation{Instituci\'o Catalana de Recerca i Estudis Avan\c{c}ats (ICREA),\protect\\ Pg.~Lluis Companys, 23, 08001 Barcelona, Spain}
\affiliation{Grup d'Informaci\'{o} Qu\`{a}ntica, Departament de F\'{\i}sica,\protect\\ Universitat Aut\`{o}noma de Barcelona, 08193 Bellaterra (Barcelona), Spain}
\orcid{0000-0001-6344-4870. }
\maketitle

\begin{abstract}
  A programmable quantum processor uses the states of a program register to specify one element of a set of quantum channels {which is applied} to an input register. It is well-known that {such a device} is impossible with {a} finite-dimensional program register for any set that contains infinitely
many unitary quantum channels (Nielsen and Chuang's \emph{No-Programming
Theorem}), meaning that a universal programmable quantum processor {does not} exist.
The situation changes if the system has symmetries. Indeed, here 
we consider group-covariant channels. If the group acts irreducibly on the channel input, {these channels} can be implemented 
exactly by a programmable quantum processor with finite program 
dimension (via 
\emph{teleportation simulation}, which uses the Choi-Jamio{\l}kowski state of the 
channel as a program). Moreover, by leveraging the representation 
theory of the symmetry group action, we show how to remove redundancy 
in the program and prove that the resulting program register has 
minimum Hilbert space dimension. 
Furthermore, we provide {upper and} lower bounds {on} the program {register}
dimension of a processor implementing all group-covariant channels 
approximately.

\end{abstract}

\section{Introduction}

Programmable quantum processors are devices which can apply desired quantum operations, specified by the user via program states, to arbitrary input states. This is convenient because one machine can be used to implement several operations and it resembles very much the way classical computers work based on classical programs and input data. Therefore, quantum processors have been studied since the early days of quantum information theory. In 1997, Nielsen and Chuang proved the \textit{No-Programming Theorem} which states that it is not possible to implement infinitely many unitary channels exactly with finite-dimensional program register~\cite{Nielsen97}, i.e.\ exact universal programmable quantum processors are impossible. However, exact programmable quantum processors were studied for special families of quantum operations \cite{HilleryZimanBusek02} and approximate programmable quantum processors were considered {in several contexts} \cite{VidalMasanesCirac02, HilleryBusekZiman02, DArianoPerinotti05, HilleryZimanBusek06, PerezGarcia06, Ishizaka08, Pirandola19, Banchi20}.
The question of optimal {program dimension} (the dimension of the program register) for an approximate universal quantum processor was only recently answered. {T}he works~\cite{Kubicki19} and~\cite{Renner20} provided new upper and lower bounds {on} the program dimension applying methods from Banach space theory and quantum entropies, respectively. 
In this work, we consider a special class
of channels, the covariant channels, i.e.\ a programmable quantum processor that implements all channels which are covariant with respect to 
representations $U$, $V$ of a compact Lie group. Note that, in particular, all results also hold for finite groups. Symmetries are of fundamental importance in physics, since they give rise to conserved quantities via Noether's theorem \cite{Noether1918}. In open systems, these symmetries arise as covariant quantum channels and are studied using tools from quantum information theory \cite{Marvian2014, Cirstoiu2020}. 
From a practical point of view, symmetries often simplify problems and break the curse of dimensionality, thus making them amenable to rigorous analysis. For these reasons, covariant quantum channels appear in many different settings, such as channel discrimination, capacities and communication tasks (see Ref.~\cite{Datta17} and the references therein). Since this is a special 
set, the question arises whether those channels can be {implemented exactly} by a programmable quantum processor. In this article, we show that an exact implementation is possible if the group acts irreducibly on the channel input. While we prove upper bounds on the program dimension for general representations $U$, we focus on the case in which $U$ is irreducible. As any representation can be decomposed into irreducible ones, it is natural to start studying this scenario before investigating more general representations in future work. \\
In Section~\ref{sec:preliminaries}, we present {some} preliminaries and our notation. We consider exact programmability of group-covariant channels in Section~\ref{sec:exactprogrammability}, where we first look at a method 
based on extreme points in Subsection~\ref{sec:extremepoints}. We show that processors have a particularly simple measure-and-prepare form if and only if the commutant of the tensor representation is abelian. 
{This yields} a program dimension equal to the number of irreducible representations occurring in the direct sum decomposition of the tensor representation in 
Corollary \ref{cor:abelian}.
{Subsection}~\ref{sec:adjointrepresentation} discusses the structure of the commutant of the tensor representation in more detail. We give a different construction of covariant programmable quantum processors based on teleportation in Subsection~\ref{sec:teleportation}. This construction is subsequently concatenated with a compression map which allows us to utilize the special structure of the Choi-Jamio\l kowski states corresponding to the covariant channels.  
This leads to Theorem~\ref{thm:teleportation-upper}, where we show that we obtain a program dimension of at 
{most} the sum of the dimensions of the blocks occurring in the structure of the Choi-Jamio\l kowski states. After the analysis of exact programmability, we consider an approximate version thereof (see Section~\ref{sec:approximateprogrammability}). First, we provide approximate upper bounds in the case of arbitrary representations $U$, $V$ in Proposition \ref{upperbound}. They are in general worse than the exact bounds in Theorem \ref{thm:teleportation-upper}, but they apply more generally. This result is the only one in which we consider arbitrary representations $U$ instead of irreducible ones.
In Theorem~\ref{lowerbounds}, we provide lower bounds on the program dimension of approximate covariant quantum processors. In particular, this shows that the construction in Theorem~\ref{thm:teleportation-upper} is optimal for the exact case.

\section{Preliminaries}
\label{sec:preliminaries}
We use the following notation: Let $d_1$, $d_2 \in \mathbb N$. We denote the set of bounded linear operators $\cH_1 \to \cH_2$ with $d_1$- and $d_2$-dimensional Hilbert spaces $\cH_1$ and $\cH_2$ by $\cB(\cH_1)$ and $\cB(\cH_2)$, respectively. The set of all $d_1$-dimensional density operators is
\begin{equation}
\cD (\cH_1) = \{ \rho \in \cB(\cH_1)~|~\rho \geq 0 , \tr(\rho) =1\},
\end{equation}
analogously for $d_2$. The set of all pure states is denoted by $\cD_P$.
A quantum channel is a completely positive trace-preserving map $T : \cB(\cH_{1}) \to \cB(\cH_{2})$, where $d_1$, $d_2 \in \mathbb N$. We write $\mathrm{CPTP}(\cH_{1}, \cH_{2})$ for this set, and $\mathrm{CPTP}(\cH)$ if $d_1 = d_2 = d$.

We review some basic results from representation theory which we need in our analysis. For further details we refer to Ref.~\cite{Simon}. Throughout the paper we shortly write compact group instead of compact Lie groups.
\begin{definition}[Unitary representation]
Let $G$ be a compact group. A \textit{unitary representation} of $G$ is a continuous homomorphism from $G$ to the unitary operators $\cU_1 \assign \cU (\cH_1)$ on some complex, $d_1$-dimensional Hilbert space $\cH_1$.
\end{definition}
Since all representations in this paper will be unitary, we will refer to them as representations for brevity. Any representation can be decomposed {into irreducible representations,} which are its fundamental building blocks because those cannot be decomposed any further.
\begin{definition}[Irreducible representation {\cite[p.~157]{Simon}}]
A unitary representation $U$ of a group $G$ on a finite-dimensional vector space $\cH_1$ is called irreducible representation (irrep) if and only if the only invariant subspaces of $\{ U_g~|~g \in G\}$ are $\{0\}$ and the whole space.
\end{definition}
A fundamental result in representation theory states that we can decompose any unitary representation of a compact group into a direct sum of irreducible representations.
\begin{lemma}
[Direct sum decomposition{~\cite[Theorem VII.9.3.]{Simon}}]\mylabel{directsum}
Every representation of a compact group $G$ is equivalent to a direct sum of irreps.
\end{lemma}
\noindent Irreducible representations of abelian groups have a particularly easy form:
\begin{corollary}[Degree irrep abelian group~{\cite[Corollary II.4.3.]{Simon}}]\label{cor:abelian-simon}
If $G$ is abelian, every irrep has dimension $d = 1$.
\end{corollary}

The image of the representation, i.e.\ the set of unitaries $U_g$, $g\in G$, generates a matrix algebra (i.e.\ a finite-dimensional unital $\ast$-algebra of linear operators).
To study symmetries in the next chapter, we define the commutant of this algebra as:
\begin{definition}[Commutant]
Let $\cA$ be a matrix algebra on the Hilbert space $\cH_d$. Its commutant is
\begin{equation}
\cA^\prime \assign \{ B~|~BA = AB \ \forall A \in \cA\}.
\end{equation}
\end{definition}

Note that the commutant is again a matrix algebra. If the algebra has a special structure, its commutant also has a corresponding structure which directly follows from the structure of $\cA$ and the definition of a commutant.

\begin{lemma}[Structure of the commutant{~\cite[Thm IX.11.2.]{Simon}}]\mylabel{structurecommutant}
Let $U$ be a unitary representation of a compact group $G$ on $\cH_1$, which can be written as $\cH_1 = \overset{K}{\underset{k=1}{\oplus}} (\cH_{k} \otimes \cH_{k}^\prime)$ such that $U_g =\overset{K}{\underset{k=1}{\oplus}} U_g^{(k)} \otimes \mathds{1}_{n_k}$ for all $g \in G$ where $U^{(k)}$, $k\in \{1, \ldots, K\}$, are irreps of $G$. Furthermore, let $\cA(U)$ be the operator algebra generated by the $\{ U_g \}_{g\in G}$ and $\cA^\prime(U)$ the corresponding commutant. Then,
\begin{equation}
\begin{split}
\cA(U) &= \bigoplus_{k=1}^K \cB(\cH_{k}) \otimes \mathds{1}_{n_k},\\
\cA (U)^\prime &= \bigoplus_{k = 1}^K \mathds{1}_{b_k} \otimes \cB(\cH_{k}^\prime) .
\end{split}
\end{equation}
\end{lemma}

\begin{figure}[ht]
\centering
\includegraphics[width=1.0\textwidth]{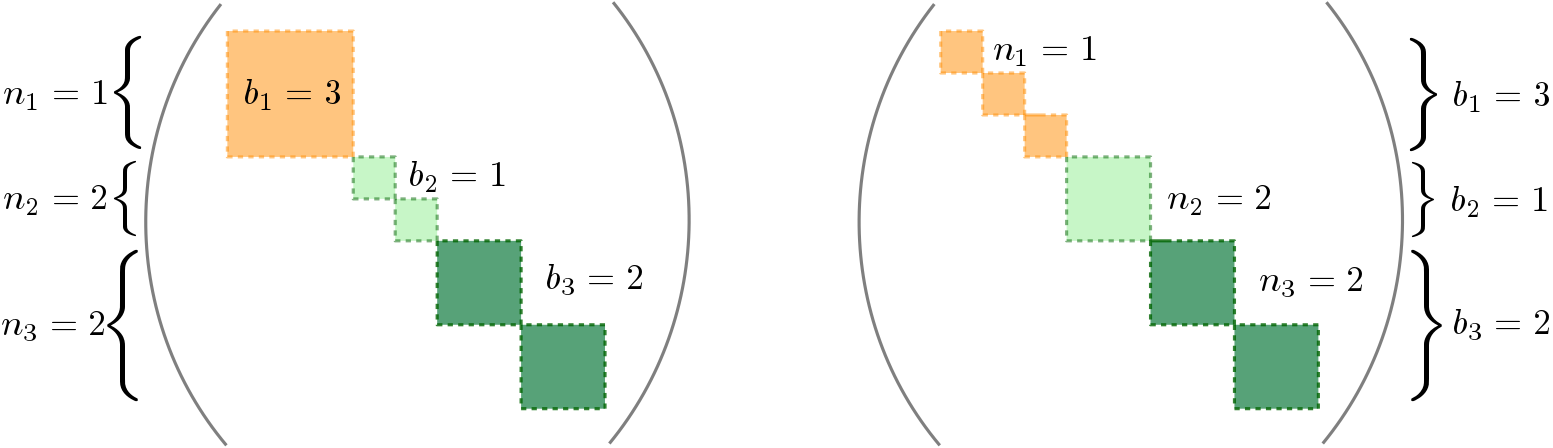}
\caption{ This figure illustrates the structure of the commutant in the case of $K=3$, i.e.\ $3$ irreps in the direct sum decomposition.}\label{fig:commutant}
\end{figure}

The special structure of the commutant is illustrated in Figure~\ref{fig:commutant} for a concrete example.

A useful tool in the analysis of direct sum decompositions, irreps and their multiplicities is the 
{notion} of a \textit{character} which is a complex number associated to each group element and related to a unitary representation through the trace operation. It is defined as{:}
\begin{definition}[Character~{\cite[p.~41]{Simon}}]
Let $U: G \to GL (\cH_1)$ be a representation such that $g \mapsto U_g$. The character $\chi: G \to \mathbb{C}$ maps every element of $G$ to a complex number according to $g \mapsto \tr(U_g)$.
\end{definition}
In the case of a one-dimensional representation, the character is equal to the one-dimensional representation. The following theorem shows how characters can be used to study representations.
\begin{lemma}[Orthogonality relations for characters {\cite[Theorem VII.9.5]{Simon}}]\label{thm:scalarproductcharacters}
For all $\alpha, \beta$ irreps,
\begin{equation}
    \lan \chi_\alpha , \chi_\beta \ran \assign \int_G \overline{\chi_\alpha (g)} \chi_\beta (g) d\mu(g) = \delta_{\alpha \beta}
\end{equation}
where $\lan \chi_\alpha , \chi_\beta \ran$ is the inner product on $L^2(G) = \{ \chi: G \to \mathbb{C} | \int_G |\chi(g)|^2 d\mu (g) < \infty \}$ and $\mu(g)$ denotes the Haar measure on $G$.
\end{lemma}
The multiplicities of irreps in the direct sum decomposition of a representation can be calculated as follows:

\begin{corollary}[Multiplicity relation~{\cite[Corollary VII.9.6.]{Simon}}]\label{cor:multiplicitycharacter}
Let $U$ be a representation of a compact Lie group $G$ and $\chi_U$ the corresponding character. Let $\chi_\alpha$ be the character of an irrep $\alpha$. {Then,}
\begin{equation}
    n_\alpha = \lan \chi_\alpha , \chi_U \ran = \int_G \overline{\chi_\alpha (g)} \chi_U (g) d\mu(g)
\end{equation}
is the multiplicity of $\alpha$ in the direct sum decomposition of $U$. Note that the $n_\alpha$ are uniquely determined by $U$.
\end{corollary}

\begin{definition}[$UV$-covariant quantum channel]
Let $G$ be a compact group 
and let $U$ and $V$ be 
{representations} on Hilbert spaces $\cH_1$ and $\cH_2$. 
Let $T: \cB (\cH_1) \to \cB (\cH_2)$ be a quantum channel. We call $T$ $UV$-covariant if 
\begin{equation}
T(U_g A U_g^\ast)= V_g T(A) {V_g}^\ast \qquad \forall A \in \cB(\cH_1),~  \forall g\in G. 
\end{equation} 
\end{definition}
\noindent The form of such channels was studied in Ref.~\cite{Datta17} for the case that $U$ is irreducible, $V = U$ and $\bar U \otimes U$ multiplicity-free. While we will allow for general $V$, most of this article is concerned with the case in which $U$ is irreducible.

\noindent The set of all $UV$-covariant channels is represented by 
\begin{equation}
\begin{split}
\cT_{UV} \coloneqq \{ T:  \cB (\cH_1) \to \cB (\cH_2) , \ &UV\text{-covariant quantum channel} \}.\\
\end{split}
\end{equation}

It is standard in quantum information theory to identify quantum channels with positive matrices via the Choi-Jamio\l kowski isomorphism~\cite{Choi75, Jamiolkowski72}. We refer to Ref.~\cite{Heinosaari} for a good book on quantum information theory. Let {$d_1 = d_2 = d$ and} $\ket{\Omega} = 1/\sqrt{d} \sum_{i = 1}^d \ket{i} \otimes \ket{i}$ be {a} maximally entangled state, where $\{\ket{i}\}_{i \in \{1, \ldots, d\}}$ {is an orthonormal basis on $\mathcal H_1$ and $\mathcal H_2$, respectively.} We define the set of all Choi-Jamio\l kowski states corresponding to {quantum channels} $T \in \cT_{UV}$ as
\begin{equation}\label{def:setc_T}
\cJ_{UV} \assign \{ c_T \in \cB(\cH_1 \otimes \cH_2) : c_T \assign (\mathrm{id} \otimes T) (\ket \Omega \bra \Omega ) ~\forall T \in \cT_{UV} \}.
\end{equation}
\noindent The $UV$-covariance property on channel level has a well-known correspondance on state-space level which is stated in the following lemma.
\begin{lemma}\mylabel{lemmaone}
The covariance property of a channel $T \in \cT_{UV}$ w.r.t.\ the unitary representations $U$, $V$ of a group $G$ is equivalent to the condition that the corresponding Choi-Jamio\l kowski state $c_T \in \cJ_{UV}$ commutes with $ \bar U_g \otimes V_g $, i.e.\ $[c_T, \bar U_g \otimes V_g]=0$ for all $g \in G$.
\end{lemma}

\begin{proof}
Using $(\bar U_g \otimes U_g) \ket \Omega \bra \Omega (\bar U_g \otimes U_g)^\ast = \ket \Omega \bra \Omega$, we can compute that 
\begin{equation}
\begin{split}
(\bar U_g \otimes V_g)^* c_T (\bar U_g \otimes V_g) 
&=(\bar U_g \otimes V_g)^*(\mathrm{id} \otimes T) (\ket \Omega \bra \Omega) (\bar U_g \otimes V_g)\\
&=(\bar U_g \otimes V_g)^*  ( \mathrm{id} \otimes T) ((\bar U_g\otimes U_g)\ket \Omega \bra \Omega( \bar U_g \otimes U_g)^* )  (\bar U_g \otimes V_g) \\ 
&= c_{\tilde T},
\end{split}
\end{equation}
where $\tilde T(\cdot) = V_g^\ast T ( U_g \cdot U_g^\ast ) V_g$.
Since the Choi-Jamio\l kowski map $T\mapsto c_T$ is an isomorphism, it follows that that the $UV$-covariance property of $T$ is equivalent to the corresponding Choi-Jamio\l kowski state commuting with $\bar U_g \otimes V_g$.
\end{proof}

The \emph{No-Programming Theorem} states that it is not possible to build a device which can implement all unitary channels, or in fact any infinite 
set of unitaries, exactly and with a finite-dimensional program register.
In this work, we consider a setting where the No-Pogramming Theorem is not applicable because we do not want our processor to implement all unitary channels but a family with a certain symmetry consisting of possibly noisy quantum operations. Therefore, we study exact programmability first before considering an approximate version.

\section{Exact and Approximate Programmability}\label{sec:exactprogrammability}

We consider a programmable quantum processor that implements all $UV$-covariant channels 
for a unitary representation $U$ of a compact group $G$.

We define a programmable quantum processor with input $\rho \in \cD(\cH_1)$ that implements all covariant channels $T \in \cT_{UV}$ with program states ${\pi_T} \in \cD(\cH_P)$ of dimension $d_P$. This is schematically illustrated in Figure~\ref{fig:processor}. Mathematically, we define this processor as follows:

\begin{figure}[ht]
\centering
\includegraphics[width=0.6\textwidth]{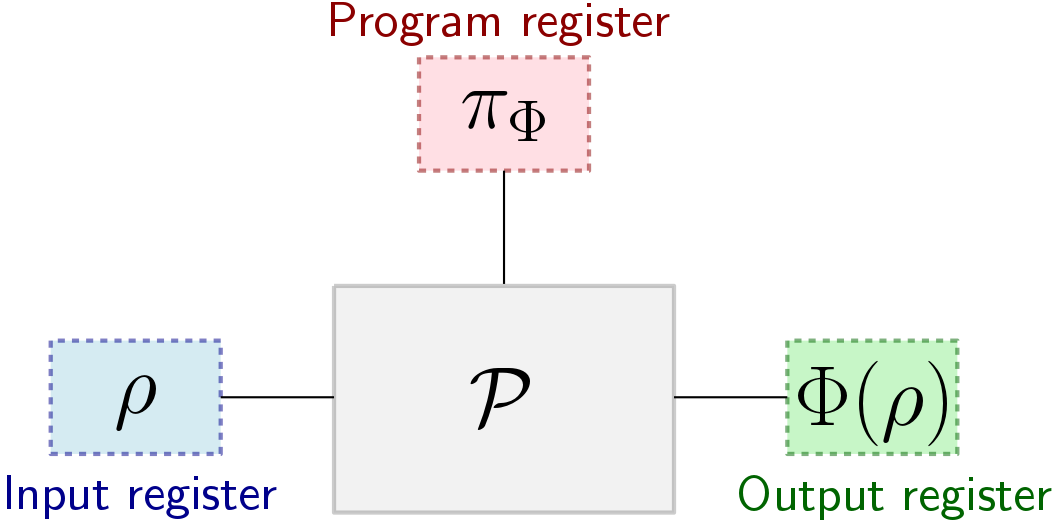}
\caption{The figure shows a CPQP$_{UV}$ with its input, program and output register. }\label{fig:processor}
\end{figure}

\begin{definition}[{$\epsilon\text{-PQP}_{\cC}$}]
\label{def:epsilonCPQP}
Let $\cH_1$ and $\cH_2$ be separable Hilbert spaces. Then we call $\cP \in \text{CPTP}(\cH_1 \otimes \cH_P,\cH_2)$, with finite-dimensional $\cH_P$, an \textit{$\epsilon$-programmable quantum processor for a set $\cC \subset \text{CPTP}(\cH_1,\cH_2)$} of channels ($\epsilon$-PQP$_{\cC}$), if for every {quantum channel} $T \in \cC$ there exists a state $\pi_T \in \cD(\cH_P)$ such that
\begin{equation}
\frac12 \| \cP(\cdot \otimes \pi_T)
           - T(\cdot) \|_\diamond \leq \epsilon.
\end{equation}
To address the Hilbert spaces $\cH_1$, $\cH_2$ and $\cH_P$, we refer to the former two as the \textit{input} and \textit{output registers} and to the latter as \textit{program register}. We say that the processor $\cP$ $\epsilon$-\emph{implements} the class $\cC$ of channels; for 
$\epsilon=0$ we say that $\cP$ exactly implements the class $\cC$, 
 and address it as a $\text{PQP}_\cC$.
\end{definition}

In particular, when $\cC = \cT_{UV}$, we write CPQP$_{UV}$ for the {covariant programmable quantum processor} and $\epsilon$-CPQP$_{UV}$ for the approximate version thereof. 
To allow for potentially mixed states in the program register is natural, since the set $\mathcal T_{UV}$ is convex, whereas the set of pure states is not.

\begin{remark}
\label{rem:pure-mixed}
In the literature, where the set of channels $\cC$ often consists 
of isometries or even unitaries, it is customary to impose that 
the program state $\pi_T$ is pure. While this does not affect the 
upper bounds on the program dimension (allowing mixed states cannot  
make it more difficult to program a processor compared to pure 
states), it is in particular an essential assumption for some lower 
bounds when unitaries are implemented~\cite{Kubicki19, Renner20}. On the other hand, it does not affect our lower bounds 
based on the properties of the Holevo information.

It is however always possible to modify a given processor $\cP 
\in \text{CPTP}(\cH_1\otimes\cH_P,\cH_2)$ to a new processor 
$\cP' \in \text{CPTP}(\cH_1\otimes\cH_P',\cH_2)$ such that every 
channel implemented by $\cP$, i.e.\ $T^\epsilon=\cP(\cdot\otimes\pi_T)$, 
is implemented by $\cP'$ using a pure program state, i.e.\ $T^\epsilon=\cP'(\cdot\otimes\proj{\psi_T})$ for $\ket{\psi_T}\in\cH_P^\prime$. 
The idea is to purify $\pi_T$ on a larger system, which gives 
rise to two upper bounds on $d_P' = \dim \cH_P'$. To obtain the first, 
every state $\pi_T$ on $\cH_P$ can be purified to a pure state 
$\ket{\psi_T} \in \cH_P\otimes\cH_P =: \cH_P'$ and we can let 
$\cP' := \cP \circ (\mathrm{id}_{\cH_1 \otimes \cH_P}\otimes\tr_{\cH_P})$, showing that 
$d_P' \leq d_P^2$. 
The second upper bound is a refinement of the first, starting from the observation 
that we can decompose a mixed program state into a convex combination 
of pure states, $\pi_T = \sum_{i} p_i \proj{\psi_{T_i}}$, where each 
$\proj{\psi_{T_i}}$ implements a channel 
$T_i = \cP(\cdot\otimes\proj{\psi_T})$, and hence $T^\epsilon = \sum_i p_i T_i$ 
is a convex combination of channels. 
The quantum channels $\text{CPTP}(\cH_1,\cH_2)$ form a convex set 
in a space of dimension $d_1^2(d_2^2-1)$, which can be seen from the 
Choi-Jamio{\l}kowski isomorphism, hence by Carath{\'e}odory's theorem 
$T$ can be written as a convex combination of not more than 
$D := d_1^2(d_2^2-1)+1$ of the $T_i$, i.e.\ $T^\epsilon = \sum_{j=1}^D q_j T_{i_j}$, 
with $q_j\geq 0$ summing to $1$. 
Thus, the state $\pi_T' := \sum_{j=1}^D q_j \proj{\psi_{T_i}}$, 
which has rank $\leq r := \min\{D,d_P\}$, implements $T^\epsilon$ as well:
$T^\epsilon=\cP(\cdot\otimes\pi_T')$. But now we only need an $r$-dimensional 
Hilbert space $\cH_R$ to purify $\pi_T' = \tr_{\cH_R} \proj{\psi_T}$, 
$\ket{\psi_T} \in \cH_P\otimes\cH_R$. As before, we can let the 
processor be $\cP' := \cP \circ (\mathrm{id}_{\cH_1\otimes \cH_P}\otimes\tr_{\cH_R})$, 
showing $d_P' \leq d_P\,\min\{D,d_P\}$. We conclude by noting that the first method is good if $d_P$ is already small, whereas the second is preferred if $d_P$ is large. The former is the case if we consider unitarily covariant channels (see Example \ref{ex:unitarilycovariant}). The latter is the case if $\mathcal C = \mathcal T_{UV}$ and the symmetries we consider are trivial, as $d_P$ is exponentially large in the dimension $d = d_1 = d_2$~\cite{Kubicki19, Renner20}.
\end{remark}

\noindent We start with the observation that the processor can without loss of generality be chosen to be $(U \otimes \mathds 1_{d_P})V$-covariant as well:
\begin{proposition}
Let $\cP$ be an $\epsilon$-CPQP$_{UV}$. Then, there exists an $\epsilon$-CPQP$_{UV}$ $\cP^\prime$ with the same program dimension {$d_P$} that is $(U \otimes \mathds 1_{d_P})V$-covariant. 
\end{proposition}
\begin{proof}
We can construct the desired processor via twirling, i.e.\
\begin{equation*}
    \mathcal P^\prime(A) = \int_G V_g^\ast \mathcal P[( U_g \otimes \mathds 1_{d_P})A( U_g^\ast \otimes \mathds 1_{d_P})] V_g d\mu(g) \qquad \forall A \in \mathcal B(\mathcal H_{{1}} \otimes \mathcal H_{{P}}),
\end{equation*}
where $\mu$ is the Haar measure on $G$. We compute
\begin{equation*}
        \mathcal P^\prime(\rho \otimes {\pi_T}) = \int_G V_g^\ast T^{\epsilon}(U_g \rho U_g^\ast) V_g d\mu(g) 
\end{equation*}
for $T \in \mathcal T_{UV}$ and $T^\epsilon$ a quantum channel such that $\frac{1}{2}\|T - T^\epsilon\|_\diamond \leq \epsilon$, since $\mathcal P$ is an $\epsilon$-CPQP$_{UV}$. Since 
\begin{equation*}
    \int_G V_g^\ast T(U_g \rho U_g^\ast) V_g d\mu(g) = T(\rho) 
\end{equation*}
for $T \in \mathcal T_{UV}$ by covariance, it holds that 
\begin{equation*}
    \frac{1}{2}\|\mathcal P^\prime(\cdot \otimes {\pi_T}) - T\|_\diamond \leq \epsilon
\end{equation*}
as well. This shows that $\mathcal P^\prime$ is also an $\epsilon$-CPQP$_{UV}$ with the same program dimension. Using the invariance of the Haar measure, it can be verified that for any $g^\prime \in G$,
\begin{equation*}
    V_{g^\prime} \mathcal P^\prime(A) V_{g^\prime}^\ast = \mathcal P^\prime[(U_{g^\prime} \otimes \mathds 1_{d_P}) A (U_{g^\prime}^\ast \otimes \mathds 1_{d_P})] \qquad \forall A \in \mathcal B(\mathcal H_{{1}} \otimes \mathcal H_{{P}}),
\end{equation*}
which shows that $\mathcal P^\prime$ is $(U \otimes \mathds 1_{d_P} )V$-covariant as desired. 
\end{proof}

Recall Lemma~\ref{lemmaone} which states that $T \in \cT_{UV}$ is equivalent to $[c_T, \bar U_g \otimes V_g]=0$ for all $g \in G$. Due to this correspondence, we consider representations of the form $\bar U \otimes V$ (which are isomorphic to the adjoint representation of $G$ if $V = U$) with $U_g \in \cU_1$, $g \in G$ and the commutant
\begin{align} 
\mathcal K \assign \mathcal A(\bar U \otimes V)^\prime &= \{ X \in \cB( \cH_1 \otimes \cH_2) ~|~ [X, \bar U_g \otimes V_g] =0 \ \forall g \in G \}\\
&= \bigoplus_{k=1}^K\mathds{1}_{b_k} \otimes \cB (\cH_k^\prime).
\end{align}
Note that since $\cH_1$ and $\cH_2$ are finite dimensional, only $K \leq d_1d_2$ irreps can appear in the commutant with multiplicity $n_k > 0$, where $n_k = \dim(\cH_k^\prime)$. We identify these elements with an index $k \in \{1, \ldots, K\}$ motivated by the fact that we want to relate the irreps occuring in the direct sum decomposition of $\bar U \otimes V$ with the number of extreme points of $\cJ_{UV}$, for instance.

While it is not true in general that all states in $\cK$ are Choi-Jamio\l kowski states of a {quantum channel}, this is true if $U$ is an irrep. This is proven in the following lemma which aligns with results in Refs.~\cite[p.~6]{Fannes04} and~\cite[p.~7]{SU2}.

\begin{lemma}\label{lemmatwo}
Let $\mathcal K$ be as defined above and
let $U$ be an irrep of a compact group G on $\cH_{1}$. Let $V$ be a representation of $G$ on $\mathcal H_{2}$. Then $\mathcal K \cap \cD (\cH_{1} \otimes \cH_{2}) = \cJ_{UV}$. Moreover, if $V$ is an irrep, any $T \in \mathcal T_{UV}$ is unital.
\end{lemma}

\begin{proof}
``$\supseteq$''
Let $c_T \in \cJ_{UV}$. Since all elements of $\cJ_{UV}$ are Choi-Jamio\l kowski states corresponding to $UV$-covariant channels, they satisfy $c_T \geq 0$ and $\tr(c_T)=1$ by definition. Hence, $c_T \in \cD(\cH_{1} \otimes \cH_{2})$. According to Lemma~\ref{lemmaone}, $T \in \cT_{UV}$ corresponds to $[c_T , \bar U_g \otimes V_g]=0$ for all $g \in G$ and thus, $c_T\in \mathcal K \cap \cD(\cH_1 \otimes \cH_2)$.

``$\subseteq$'' 
Let us refer to 
$\mathcal H_1$ as system $A$ and 
to $\mathcal H_2$ as system $B$. If we intersect $\mathcal K$ with the set of states $\cD(\cH_1 \otimes \cH_2)$, then every $\rho_{AB} \in \mathcal K \cap \cD(\cH_1 \otimes \cH_2)$ satisfies $\tr(\rho_{AB}) =1$ and $\rho_{AB} \geq 0$ as well as $[\rho_{AB} , \bar U_g \otimes V_g ]=0$ for all $g \in G$. To obtain $\rho_{AB} \in \cJ_{UV}$, we additionally have to show the required property $\tr_{B}(\rho_{AB}) = \frac{\mathds{1}_{d_1}}{d_1}$. 
Using $[\rho_{AB}, \bar U_g \otimes V_g]=0$, we get
\begin{equation}
\tr_B(\rho_{AB})= \tr_B\big(( \bar U_g \otimes V_g ) \rho_{AB} (\bar U_g \otimes V_g)^* \big)= \bar U_g \tr_B(\rho_{AB}) \bar U_g^*.
\end{equation}
for any $g \in G$ which is equal to
\begin{equation}
\tr_B( \rho_{AB}) \bar U_g = \bar U_g \tr_B (\rho_{AB}).
\end{equation} 
Since $U$ is an irrep if and only if $\bar U$ is, we infer due to Schur's Lemma:
\begin{equation}
\tr_B(\rho_{AB}) = \lambda \cdot \mathds{1}_{d_1} \qquad \textrm{for some} \ \lambda  \in \mathbb{C}.
\end{equation}
Taking the trace on both sides results in
\begin{align}
1 = \tr \big( \tr_B (\rho_{AB}) \big) = \lambda \cdot \tr(\mathds 1_{d_1}) = \lambda  \cdot d_1.
\end{align}

Hence, $\lambda = 1/d_1$.
This yields $\tr_B (\rho_{AB}) = \frac{\mathds{1}_{d_1}}{d_1}$ which we aimed to show. With the same reasoning, we can also conclude that if $V$ is an irrep,
\begin{equation}\mylabel{unitalproperty}
\tr_A (c_T) = \frac{\mathds{1}_{d_2}}{d_2}
\end{equation}
for all $c_T \in \cJ_{UV}$, which implies that $T$ is unital.
\end{proof}

In Ref.~\cite{HilleryZimanBusek02}, the authors derived that channels implemented by a processor that is covariant with respect to the special unitary group $SU(\cH_1)$ are unital using a similar argument.
We will now consider how to construct covariant programmable quantum processors in the case where $\mathcal K$ is abelian.

\subsection{Covariant programmable quantum processors from extreme points}\label{sec:extremepoints}

In this section, we consider a special class of CPQP$_{UV}$, the measure-and-prepare CPQP$_{UV}$. We will show that $\mathcal T_{UV}$ can be implemented by a  measure-and-prepare CPQP$_{UV}$ if and only if the commutant $\mathcal K$
 is abelian.
Before we state the main result, we present a lemma required for its proof which is a correspondence between the commutant and the affiliated state space which we require in order to show our first main result for exact programmability.

\begin{lemma}\mylabel{lemmathree}
Let $U$ be an irrep of a compact group $G$, and let $V$ be another representation of $G$. Then, the following are equivalent:
\begin{enumerate}
    \item $\mathcal K$ is abelian.
    \item $\cJ_{UV}$ is isomorphic to a simplex.
    \item $\cJ_{UV}$ is isomorphic to a polytope.
\end{enumerate}

\end{lemma}

\begin{proof}
``i) $\Rightarrow$ ii)''  
This statement is mentioned in Ref.~\cite{VollbrechtWerner01} without proof. Let $B \in \cK$ and let $\mathcal K$ be an abelian
{matrix} algebra. This implies that $n_k=1$ for all $k \in \{1, \ldots, K\}$. Furthermore, let $K$ be the number of irreps appearing in the direct sum decomposition of $\bar U\otimes V$.  
We obtain
\begin{equation}
B= \bigoplus_{k=1}^K  \mathds{1}_{b_k} \otimes x_k 
\end{equation}
where $x_{k} \in \mathbb{C}$. 
Moreover, $B \geq 0$ if and only if $x_k \geq 0$ for all $k \in \{1, \ldots, K\}$ and $\tr(B)=1$ if and only if $\sum_k b_k x_k =1$.
According to Lemma~\ref{lemmatwo}, extreme points of $\mathcal K \cap \mathcal D(\mathcal H_1 \otimes \mathcal H_2)$ 
are also extreme points of $\cJ_{UV}$. 
The extreme points of $\mathcal K \cap \mathcal D(\mathcal H_1 \otimes \mathcal H_2)$ are of the form  $x_i = 1/b_i$ for some $i \in \{1, \ldots, K\}$ and $x_l = 0$ for all $l \neq i$.
We identify the $K$ extreme points of $\cJ_{UV}$ with $K$ points in $\mathbb{R}^K$. Therefore, $\cJ_{UV}$ is isomorphic to a $(K-1)$-simplex.

``iii) $\Rightarrow$ i)'' 
We prove this statement by contraposition, i.e.\ we show if $\mathcal K$ is non-abelian, $\cJ_{UV}$ is not isomorphic to a polytope. If $\mathcal K$ is non-abelian, there is a $k \in \{1, \ldots, K\}$ such that the corresponding 
block is of dimension $n_k > 1$. 
Let us consider elements of the form
\begin{equation}\mylabel{purestatestructure}
B_{{\varphi}}=\bigg( \frac{1}{b_k} \ket{\varphi_k} \bra{\varphi_k} \otimes \mathds{1}_{b_k}\bigg) \oplus \textbf{0},
\end{equation} 
where $\ket{\varphi_k} \bra{\varphi_k}  \in \mathcal D(\mathcal H_{k}^\prime)$. The normalization $\frac{1}{b_k}$ yields $\tr(B_{{\varphi}})=1$ and furthermore, $B_{{\varphi}} \geq 0$, i.e.\ $B_{{\varphi}}  \in {\mathcal D}(\mathcal H_1 \otimes \mathcal H_2)$. 
These elements are extreme points of $\cJ_{UV}$ by Lemma~\ref{lemmatwo}. 
Thus, there are infinitely many extreme points of $\cJ_{UV}$. Hence, the set cannot be isomorphic to a polytope which has finitely many extreme points by definition.

The last implication ``ii) $\Rightarrow$ iii)'' is obvious.
\end{proof}

Let us now state the definition of the class of processors we consider in this section.

\begin{definition}[Measure-and-prepare $PQP_\mathcal C$]
Let $\mathcal H_1$ and $\mathcal H_2$ be separable Hilbert spaces. Then, we call a PQP$_{\mathcal C}$ $\mathcal P$ with finite-dimensional $\mathcal H_P$ a measure-and-prepare PQP$_\mathcal C$ for a set of quantum channels $\mathcal C$ if there exists a $K \in \mathbb N$, a POVM $\{E_i\}_{i \in \{1, \ldots, K\}} \subset \mathcal D(\mathcal  H_P)$ and a set of quantum channels $T_k \in \mathcal C$, $k \in \{1, \ldots, K\}$, such that
\begin{equation*}
    \mathcal P(A \otimes B) = \sum_{k = 1}^K \mathrm{tr}[E_k B] T_k(A) \qquad \forall A \in \mathcal B(\mathcal H_1),~B\in \mathcal B(\mathcal H_P).
\end{equation*}
\end{definition}

\begin{proposition}\label{TheoremCovariant}
Let $\mathcal C \subset \mathrm{CPTP}(\mathcal H_1, \mathcal H_2)$ be a convex set of quantum channels. Then, there is a measure-and-prepare PQP$_{\mathcal C}$ if and only if $\mathcal J_{\mathcal C}$, the corresponding set of Choi-Jamio\l kowski states, is isomorphic to a polytope. Moreover, if such a measure-and-prepare PQP$_{\mathcal C}$ exists, $d_P$ can be chosen to be the number of extreme points and the program states can be chosen pure.
\end{proposition}

\begin{proof}
We fix a quantum channel $T \in \mathcal C$ with its corresponding Choi-Jamio\l kowski state $c_T \in \mathcal J_{\mathcal C}$. 
Since $\mathcal J_{\mathcal C}$ is isomorphic to a polytope, it is spanned by $K$ extreme points {$c_{T_k}$} 
and can therefore be written as convex combination of these
\begin{equation}
{c_T= \sum_{k = 1}^K  x_k c_{T_k},}
\end{equation}
{where $x_k \in [0,1]$ and $\sum_{k = 1}^K x_k = 1$.} 
By the Choi-Jamio\l kowski isomorphism, there is a channel {$T_k \in \cT_{UV}$} corresponding to each of the extreme points {$c_{T_k}$} of $\mathcal J_{\mathcal C}$, i.e.\ $T$ can be linearly decomposed 
\begin{equation}
{T(\cdot)= \sum_{k = 1}^K x_k T_k (\cdot)}
\end{equation}
with extreme points {$T_k (\cdot)$}. We encode the {$\{x_k\}_{k \in \{1,\ldots, K\}}$} in the program state as follows 
\begin{equation}
{\ket {\psi_T}= \sum_{k = 1 }^K \sqrt{x_k} \ket k \in \cD_P(\cH_P),}
\end{equation}
with an arbitrary orthonormal basis {$\{\ket k \}_{k \in \{1, \ldots, K\}}$} on {$\cH_P$}.
The following processor implements $T \in \mathcal C$ exactly with a program register of dimension $d_P=K$:
\begin{equation*}
    {\cP (A\otimes B ) = \tr_{{\cH_P}} \left[\sum_{k = 1}^K T_k(A) \otimes\lan k| B |k \ran \ket k \bra k \right] \qquad \forall A \in \mathcal B(\mathcal H_{1}),~\forall B \in \mathcal B(\mathcal H_P) }
\end{equation*}
and extended by linearity. We verify that this is indeed a measure-and-prepare PQP$_{\mathcal C}$:
\begin{equation*}
\cP (\rho\otimes \ket{\psi_T} \bra{\psi_T} ) 
 = \sum_{k = 1}^{K}T_k(\rho) \mathrm{tr}[E_k \ket {\psi_T} \bra{\psi_T}] 
 = \sum_{k = 1}^K x_k T_k (\rho) = T(\rho),
\end{equation*}
where, $E_k = \ket{k}\bra{k}$ for all $k \in \{1, \ldots, K\}$. Thus, we showed that if $\mathcal J_{\mathcal C}$ is isomorphic to a polytope, there is a measure-and-prepare processor $\cP$ that implements all $T \in \mathcal C$ exactly with program dimension $\dim(\cH_P)= d_P = K$.

Conversely, let $\mathcal P$ be a measure-and-prepare PQP$_{\mathcal C}$ with a $K$-outcome POVM. Then, $\mathcal C$ is the convex hull of at most $K$ extreme points, since by definition for any $T \in \mathcal C$
\begin{equation*}
    T = \mathcal P(\cdot \otimes \pi_T) = \sum_{k = 1}^K p_k T_k
\end{equation*}
with $p_k = \mathrm{tr}[E_k \pi_T]$, i.e.\ $(p_1, \ldots, p_K)$ is a probability distribution. Thus, the extreme points of $\mathcal C$ are a subset of $\{T_1, \ldots, T_K\}$ and hence $\mathcal C$ is a polytope.
\end{proof}

The following corollary assures that if we want to check whether channels $T\in\cT_{UV}$ are programmable exactly by a measure-and-prepare CPQP$_{UV}$ with finite-dimensional program register, we can consider the specific structure of the commutant $\mathcal K$.

\begin{corollary} \label{cor:abelian}
Let $U$ be an irrep of a compact group $G$ {on $\cH_1$} and $V$ a representation of $G$ {on $\cH_2$}. 
Furthermore, let $\cK$ be the commutant of $\bar U\otimes V$. 
Then, $\cK$ is abelian if and only if there is a measure-and-prepare CPQP$_{UV}$ $\cP$ that implements all $T\in \cT_{UV}$ exactly with $d_P \leq K$, where $K$ is the number of irreps appearing in the direct sum decomposition of $\bar U\otimes V$.
\end{corollary}

\begin{proof}
The proof directly follows from Lemma~\ref{lemmathree} and Proposition \ref{TheoremCovariant}.
\end{proof}

We give an example of a group for which exact programmability holds with a measure-and-prepare CPQP$_{UV}$.
\begin{example}[Unitarily covariant channels]\label{ex:unitarilycovariant}
We consider a measure-and-prepare CPQP$_{UV}$ that implements all $UU$-covariant channels 
where $U$ is the defining representation of the group $G=\cU_1$. 
The tensor representation $\bar U \otimes U$ can be decomposed into a direct sum of a one-dimensional and a $(d^2-1)$-dimensional irrep both with multiplicity $1$. Thus, elements of the commutant consist of a one-dimensional block with multiplicity equal to one and a one-dimensional block with multiplicity $d^2-1$, i.e.\ in this case $K=2$, $n_1=b_1=1$, $n_2=1$ and $b_2=d^2-1$. 
Due to Lemma~\ref{lemmaone}, we know that the Choi-Jamio\l kowski state $c_T \in \mathcal J_{UU}$ 
satisfies $[c_T, \bar U_g \otimes U_g]=0$. The state space therefore has the following form: 
\begin{equation}
\mathcal K \cap \cD(\cH_1 \otimes \cH_1)=\bigg \{ \hat \alpha \frac{\mathds{1}}{d^2} + (1-\hat \alpha) \ket \Omega \bra \Omega \Big| \hat \alpha \in \Big[0,\frac{d^2}{d^2-1}\Big] \bigg \}
\end{equation}
with extreme points $\ket \Omega \bra \Omega$ and $\frac{1}{d^2-1} (\mathds{1} - \ket \Omega \bra \Omega)$. Thus, $T$ has the form $T(\cdot)=\hat\alpha \tr (\cdot) \frac{\mathds{1}}{d} +(1-\hat\alpha) \mathrm{id}$~\cite{Keyl02, VollbrechtWerner01}.
Every $c_T$  can be written as convex combination of the two extreme points and the set of all convex combinations 
\begin{equation}
\mathcal K \cap \cD(\cH_1 \otimes \cH_1)=\bigg \{ x \ket \Omega \bra \Omega + (1-x) \frac{1}{d^2-1} (\mathds{1} - \ket \Omega \bra \Omega ) ~\Big|~ x \in \big[0,1] \bigg \}
\end{equation}
is isomorphic to a $1$-simplex. Thus, the measure-and-prepare CPQP$_{UV}$ can be implemented with program dimension $d_P = 2$ using the construction in Proposition \ref{TheoremCovariant}.

\end{example}

\subsection{Structure of the commutant of the tensor representation}\label{sec:adjointrepresentation}

Based on the previous section, one could argue that in the case where the direct sum decomposition of $\bar U\otimes V$ consists of at least one one-dimensional irrep with multiplicity $n_k >1$, it would not be possible to implement the corresponding $UV$-covariant channel exactly in this case. It is instructive to see why this situation never arises, which we will discuss in this section.

The argument is the following: Assume $d_1 = d_2 = d$ and that there is a $k \in \{1, \ldots, K\}$ such that $n_k >1$. Then, there would be  
elements in $\cJ_{UV}$ of the form
\begin{equation}
B_{{\varphi}}=\ket {\varphi_k} \bra {\varphi_k} \oplus \textbf{0} \in \cJ_{UV}
\end{equation}
where $\ket {\varphi_k} \bra {\varphi_k} \in \cD(\cH_{k}^\prime)$ is a rank-one projector and $\dim(\textbf{0})=d^2-n_k$. By the Choi-Jamio\l kowski isomorphism, there is a corresponding channel $T\in \cT_{UV}$. 
 We consider a Kraus representation of the channel $T$. Since $T$ is a completely positive map and rank$(B_\varphi)=1$, the channel $T$ can be written as {$T(\cdot) = X (\cdot) X^\ast$ with one Kraus operator $X\in \cB(\cH_2)$. Since, additionally, $T$ is trace-preserving, we know that $XX^* = \mathds{1} = X^*X$}, i.e.\ the Kraus operator is a unitary and $T$ is a unitary channel
\begin{equation}
T (\cdot)= W (\cdot) W^\ast
\end{equation}
with {$W \in \cU_2$}.

Since there are infinitely many 
{pure states}, the processor would have to implement infinitely many corresponding {unitary channels.}  
This contradicts the No-Programming Theorem according to which there is no processor that implements infinitely many unitary channels 
{exactly} with program dimension $d_P<\infty$. This shows that a processor $\cP$ with $d_P< \infty$
cannot exist if $b_k = 1$ and $n_k > 1$ for some $k \in \{1, \ldots, K\}$. Note that the above argument fails for $b_k > 1$, since then there might be no rank-$1$ elements in $\cK$.

However, from a representation-theoretic perspective this situation does not arise because a one-dimensional irrep in the direct sum decomposition of $U\otimes \bar V$ always has multiplicity $\leq 1$ as the following proposition shows.
\begin{proposition}
Let $U$ be an irrep {on $\cH_1$} of a 
compact group $G$. 
Let $V$ be another representation of $G$ {on $\cH_2$ with} dimension $d_2 \leq d_1$.
In a direct sum decomposition of $\bar U \otimes V$, the one-dimensional irreps $\lambda$ appear with multiplicity $n_\lambda\leq 1$.
\end{proposition}
\begin{proof}
 
Let $\chi_U$ be the character of the irrep $U$, $\chi_V$ the character of $V$, and $\lambda$ the character of the one-dimensional irrep $\lambda$. We use the following scalar product (see Lemma~\ref{thm:scalarproductcharacters})
\begin{equation}
\lan \chi , \psi \ran = \int_{G} \overline{\chi(g)} \psi(g) d\mu (g),
\end{equation}
where $\mu$ is the Haar measure on $G$. Here, $\psi$ is the character of an arbitrary representation of $G$. Note that if $U$ is an irrep, then this scalar product of the corresponding characters gives the multiplicity of $U$ in the representation corresponding to $\psi$ (see Corollary~\ref{cor:multiplicitycharacter}). 
We want to show that the multiplicity of $\lambda$ in $\bar U \otimes V$ is $\leq 1$. Note that the character of $\bar U \otimes V$ is $\bar \chi_U \cdot \chi_V$. Let $\hat G$ be the set of irreps of $G$ and let
\begin{equation*}
    \chi_V = \sum_{\alpha \in \hat G} n_\alpha \chi_\alpha
\end{equation*}
be the decomposition of $V$ into irreducible representations on the level of characters. Then,
\begin{equation}
\begin{split}
\lan \lambda , \bar \chi_U \cdot \chi_V \ran &= \int_{G} \overline{\lambda (g)} \big( \overline{\chi_U (g)} \chi_V(g) \big) d\mu(g) \\
& = \int_{G} \overline{\lambda (g) \chi_U (g)}  \chi_V(g) d\mu(g) \\
&= \lan \lambda \cdot \chi_U , \chi_V \ran. \\
&=\sum_{\alpha \in \hat G} n_\alpha \lan \lambda \cdot \chi_U, \chi_\alpha \ran. \\
\end{split}
\end{equation}
Since $\lambda$ is the character of a one-dimensional irrep, it is equal to the representation itself and $|\lambda(g)|^2 = 1$ for all $g \in G$. Thus, $\lan \lambda \cdot \chi_U , \lambda \cdot \chi_U \ran = 1$ and the representation corresponding to $\lambda \cdot \chi_U$ is again irreducible. Note that the representation corresponding to $\lambda \cdot \chi_U$ has the same dimension as $U$.

The scalar product $\lan \lambda \cdot \chi_U , \chi_\alpha \ran$ thus gives the multiplicity of $\lambda \cdot \chi_U$ in $\chi_\alpha$. 
Let $\alpha$ be such that $n_\alpha > 0$. If the dimension of the representation corresponding to $\chi_\alpha$ is smaller than $d_2$, then $\lan \lambda \cdot \chi_U , \chi_\alpha \ran = 0$ by Lemma \ref{thm:scalarproductcharacters}. Thus, $\lan \lambda , \bar \chi_U \cdot \chi_V \ran = 0$. If the dimension is $d_2$, then $\chi_V = \chi_\alpha$ for dimensional reasons and $V$ is irreducible.
The multiplicity of an irrep in another irrep can {be at most} equal to $1$ which proves the assertion.
\end{proof}
\begin{remark}
We can also give a direct algebraic argument for the impossibility of a one-dimensional irrep of $G$ to occur in $\bar{U}\otimes V$ with multiplicity larger than one, if $U$ is 
 an irrep and $V$ is an irrep, too, and has the same dimension as $U$. Namely, observe that all pure states in the 
 multiplicity space $\cK$ of a one-dimensional irrep have to be 
 maximally entangled due to Schur's Lemma (their reduced states 
 on both factors have to be maximally mixed). However, there is no two- or higher-dimensional subspace in $d \times d$ that merely consists of maximally entangled states (see, for instance Ref.~\cite[Proposition {6}]{Cubitt08}). 
\end{remark}

After this discussion, we give a different construction of covariant programmable quantum processors which is more widely applicable.

\subsection{Covariant programmable quantum processors from teleportation}
\label{sec:teleportation}
If $\mathcal K$ is not abelian, we need a different method to show exact programmability. 
This case appears for example for the finite group $A_4$~\cite[p.~20]{FultonHarris}, which is outside the scope of Ref.~\cite{Datta17}. 

\begin{example}
The alternating group $A_4$ is a subgroup of the symmetric group $S_4$ consisting of only the even permutations: 
\begin{equation}
  A_4 = \{e,(12)(34),(13)(24),(14)(23), (123), (134), (243), (142), (321), (431), (342), (241)\}. 
\end{equation}
Note that $K_4 = \{e,(12)(34),(13)(24),(14)(23)\}$ is an abelian subgroup of $A_4$, in fact isomorphic to the Klein group, and since it can be described as the set of all elements of order at most $2$, it is a normal subgroup. Its cosets are $(123)K_4$ and $(321)K_4 = (123)^2 K_4$, which are conjugacy classes, and the quotient $A_4/K_4$ is the cyclic group $\mathbb{Z}/3$ of three elements.

The group has four irreps $\varphi^{(0)}$, $\varphi^{(1)}$, $\varphi^{(2)}$, and $\vartheta$. The first three, $\varphi^{(0)}$, $\varphi^{(1)}$, $\varphi^{(2)}$, are one-dimensional, while $\vartheta$ is three-dimensional~\cite[Section 2.3]{FultonHarris}.
Indeed, with the third root of unity $\zeta = e^{2\pi i/3}$,
we have $\varphi^{(j)}\bigl((123)^k K_4\bigr) = \zeta^{jk}$.
On the other hand, $\vartheta$ has a very intuitive form as the real $\text{SO}(3)$ symmetries of a regular tetrahedron in three-space: concretely, let 
$\cH=(1,1,1,1)^\perp = \bigl\{(a_1,a_2,a_3,a_4) : \sum_i a_i = 0\bigr\} \subset \mathbb{C}^4$, 
which is spanned by the vertices $v_1=(3,-1,-1,-1)$, $v_2=(-1,3,-1,-1)$, $v_3=(-1,-1,3,-1)$, $v_4=(-1,-1,-1,3)$ of the tetrahedron. The group action of $\vartheta$ is simply by permutation of the coordinate axes, and so all its matrices are real; as a consequence, $\overline\vartheta = \vartheta$.

Consider now the set of $\vartheta\vartheta$-covariant channels. Note that they form a semigroup (since they are closed under composition), and at the same time a closed convex set, isomorphic to the $3\times 3$-states invariant under $\vartheta \otimes \vartheta$.  The latter representation decomposes as
\begin{equation}
  \label{eq:vartheta-decomposition}
  \vartheta \otimes \vartheta \simeq \varphi^{(0)} \oplus \varphi^{(1)} \oplus \varphi^{(2)} \oplus 2 \vartheta,
\end{equation}
where $\vartheta$ appears with multiplicity $2$. Hence, there are three one-dimensional blocks on the diagonal and two three-dimensional ones, i.e.\ the commutant consists of three one-dimensional blocks and a two-dimensional block of multiplicity three. Thus, Lemma \ref{lemmathree} shows that $J_{\vartheta\vartheta}$ is not a polytope; indeed its extreme points are three maximally entangled pure states, as well as a two-dimensional family of rank-three states, parametrized by the surface of the Bloch sphere. This implies that Corollary \ref{cor:abelian} does not cover all cases of interest. 

We can give a more explicit description of the extreme points and the associated quantum channels. To start, because of the semigroup property, the unitary covariant channels form a group, and since by the preceding representation-theoretic analysis it has three elements, corresponding to the three maximally entangled pure states in $J_{\vartheta\vartheta}$, we are looking for a unitary $V$ of degree $3$ that commutes with $\vartheta$ up to a phase given by the one-dimensional irreps: $\vartheta_g V \vartheta_g^* = \varphi^{(1)}_g V$. Then, the unitary channels $T_j(\rho)=V^j \rho V^{\ast j}$ ($j=0,1,2$) are $\vartheta\vartheta$-covariant. For instance, $V$ can be given as 
\begin{alignat}{8}
  V\ket{v_1} &= \frac12\bigl({}                &&\phantom{+\zeta}\ket{v_2} &&+\zeta\ket{v_3}   &&+\zeta^2\ket{v_4}\bigr), \\
  V\ket{v_2} &= \frac12\bigl(\phantom{\zeta}\ket{v_1} &&{}                 &&+\zeta^2\ket{v_3} &&+\zeta\ket{v_4}\bigr), \\
  V\ket{v_3} &= \frac12\bigl(\zeta\ket{v_1}   &&+\zeta^2\ket{v_2} &&{}                 &&+\phantom{\zeta}\ket{v_4}\bigr), \\
  V\ket{v_4} &= \frac12\bigl(\zeta^2\ket{v_1} &&+\zeta\ket{v_2}   &&+\phantom{\zeta}\ket{v_3}         &&\phantom{+\zeta^2\ket{v_4}}\bigr),
\end{alignat}
keeping in mind that this description is redundant (but consistent), as $\ket{v_1}+\ket{v_2}+\ket{v_3}+\ket{v_4}=0$ and the sum of the r.h.s.~vectors vanishes, too. This defines a unitary because the pairwise inner product of any two of the r.h.s.~vectors equals the inner product of the corresponding $\ket{v_i}$ on the l.h.s..

The remaining non-unitary, yet unital, channels, we have to find, are those whose Choi-Jamio{\l}kowski states are extreme points of rank $3$ of $J_{\vartheta\vartheta}$, which means that the channels have Kraus rank $3$. There is one distinguished element in this set, namely the Werner-Holevo channel~\cite{WernerHolevo}
\begin{equation}
  W(\rho) = \frac12\left( (\tr\rho)\mathds{1} - \rho^T \right), 
\end{equation}
where $T$ denotes the transpose, whose Choi-Jamio{\l}kowski state $\alpha$ is the normalized projector onto the $3\times 3$-antisymmetric subspace. It is $\bar{U}U$-covariant for arbitrary $U\in\text{SU}(3)$; in particular it is $\vartheta\vartheta$-covariant if we recall that $\vartheta$ is real, but it is much more symmetric than that. It has the curious property that it is extremal in the set of all CPTP maps, hence in particular in the set of covariant ones. Moreover, we note that the channels $T_j\circ W\circ T_k$ ($j$, $k=0,1,2$) are all $\vartheta\vartheta$-covariant by the semigroup property, giving further extreme points. Their Choi-Jamio{\l}kowski states, which are $(\bar{V}^k\otimes V^j)\alpha(\bar{V}^k\otimes V^j)^*$, generate the qubit algebra $\mathcal{A}$ of the multiplicity space of $\vartheta$ in Eq.~\eqref{eq:vartheta-decomposition}. In particular, we will find subsequently elements from the algebra which are a representation of the Pauli matrices $\sigma_X$, $\sigma_Y$, $\sigma_Z$. This was inspired by the approach taken in \cite{Eggeling2001} for a different problem.
Let $P:=\mathds{1}_9 - \ket{\Omega}\bra{\Omega} - (\mathds{1}_3 \otimes V)\ket{\Omega}\bra{\Omega}(\mathds{1}_3 \otimes V^\ast) - (\mathds{1}_3 \otimes V)^2\ket{\Omega}\bra{\Omega}(\mathds{1}_3 \otimes V^\ast)^2$ be the projection onto $\mathcal A$. Using the form of $\alpha$, which is proportional to the projector onto the antisymmetric subspace, we infer that $PF \in \mathcal A$, where $F$ is the flip (or swap) operator on $\mathbb C^3 \otimes \mathbb C^3$. Furthermore, we can verify that $[P, \mathds{1}_3 \otimes V] = 0$ and  $[P, F] = 0$. Combining this with the previous finding that the Choi-Jamio{\l}kowski states of $T_j\circ W\circ T_k$ are in $\mathcal A$, we obtain four linearly independent elements of $\mathcal A$, namely $P$, $PF$, $P(V \otimes V^\ast)$ and $PF(V \otimes V^\ast)$. Since $PF$ is traceless, $P$ is orthogonal to all elements except $P(V \otimes V^\ast)$ and $PF$ is orthogonal to all elements except $PF(V \otimes V^\ast)$. Thus, we can identify $P \simeq \mathds{1}$, $V_X:=PF \simeq \sigma_X$ and make the ansatz $\sigma_Y \simeq V_Y:= a PF(V \otimes V^\ast) + b PF$, $\sigma_Z =-i\sigma_X \sigma_Y \simeq V_Z:= -i(a P(V \otimes V^\ast) + b P)$. Imposing $\mathrm{tr}(V_XV_Y) = 0$ and $\mathrm{tr}(V_Y^2)=6$ yields $a=\frac{2}{\sqrt{3}}$ and $b=\frac{1}{\sqrt{3}}$: 
\begin{align}
    \mathds{1}        \simeq V_0 &= P, \\
    \sigma_X \simeq V_X &= PF, \\
    \sigma_Y \simeq V_Y &= \frac{1}{\sqrt{3}}PF\bigl(2 V\otimes V^\ast + \mathds{1}\bigr), \\
    \sigma_Z \simeq V_Z &= \frac{1}{i\sqrt{3}}P\bigl(2 V\otimes V^\ast + \mathds{1}\bigr).
\end{align}
We can verify that indeed $V_i V_j+V_j V_i = 2\delta_{ij}P$ for $i$, $j \in \{X,Y,Z\}$. Hence, we can use a Bloch ball representation for the Choi-Jamio{\l}kowski states $J$ in $\mathcal A$: $J \in \mathcal A$ if and only if
\begin{equation*}
    J = J_{\vec{\lambda}} = \frac{1}{6}\left(P+\lambda_x V_X + \lambda_y V_Y + \lambda_z V_Z\right),
\end{equation*}
where $\vec{\lambda} = (\lambda_x, \lambda_y, \lambda_z) \in \mathbb R^3$ with $\lambda_x^2+\lambda_y^2+\lambda_z^2 \leq 1$. Via the Choi-Jamio{\l}kowski isomorphism, these describe $\vartheta\vartheta$-covariant channels ${W}_{\vec{\lambda}}$. The extremal ones are precisely the previously found ${T}_0=\operatorname{id}$, ${T}_1$ and ${T}_2$, as well as the ${W}_{\vec{\lambda}}$ with $\lambda_x^2+\lambda_y^2+\lambda_z^2=1$.
\end{example}

We can construct an exact quantum processor for covariant channels based on teleportation simulation, i.e.~the simulation of quantum channels by quantum teleportation. In the case of the Pauli group this goes back to the Refs.~\cite{Bennett93, Bennett96}. Important developments concerning the teleportation of covariant channels can be found in Refs.~\cite{Horodecki98, Chiribella09, MastersThesisMH}. 
See also the very recent Ref.~\cite{Wang20}. We know from Refs.~\cite[p.~58]{Pirandola17} and~\cite[Proposition 2]{Wilde17} that it is always possible to simulate {$UV$}-covariant channels exactly using the corresponding Choi-Jamio{\l}kowski state.

This can easily be formulated as a processor which uses the Choi-Jamio{\l}kowski state as program state and performs the teleportation protocol.
Therefore, the dimension of the program register is $d_1d_2$. 
We will make this precise in the next proposition already present in Refs.~\cite{Pirandola17, Wilde17}, which we have included here for convenience.

\begin{proposition} \label{prop:teleportation-processor}
Let $G$ be a compact group and let $U$ be an irreducible representation {on $\cH_1$.} Let $V$ be another representation of $G$ {on $\cH_2$.} Then, there is a CPQP$_{UV}$ with {program dimension} $d_P \leq d_1 d_2$.
\end{proposition}

\begin{proof}
Let $\rho \in \cD(\cH_{1})$ be the state to be teleported. In this proof, we identify the input space as system $A$, $\mathcal H_1 \simeq \mathcal H_A$, and the program as a composite system with parts $A^\prime$ and $B$, $\mathcal H_{{P}} \simeq \mathcal H_{A^\prime} \otimes \mathcal H_B =: \cH_{A^\prime B}$, where $\mathcal H_{A^\prime} \simeq \mathcal H_A$ and $\mathcal H_{B}\simeq \mathcal H_2$ is isomorphic to the output space of $T$. We will write $d_1$ for the dimension of $A$. Note that $\dim A = \dim A^\prime= \dim A^{\prime \prime} =d_1$. The Choi-Jamio\l kowski states corresponding to the channels that are simulated serve as program states of the processor running the following protocol:
\begin{enumerate}
\item the processor measures according to the POVM
\begin{equation}
\{M_g\}_{g \in G} \assign \{ d_1^2 (\mathds{1} \otimes \bar U_g ) \ket \Omega \bra \Omega_{A A^\prime} (\mathds{1} \otimes \bar U_g )^\ast d\mu(g) \}_{g \in G}.
\end{equation}
Due to Schur's Lemma this is a POVM. Note that this POVM can potentially be continuous.
\item On outcome $g$, apply $V_g^\ast (\cdot) V_g$ to the outcome of the protocol.
\end{enumerate}
This construction of a processor implements {$UV$}-covariant channels $T \in \cT_{UV}$ with the Choi-Jamio\l kowski state as program state. The map is defined as
\begin{equation}\label{def:teleportationprocessor}
\cP(X \otimes Y) \assign  \int_{G} V_g^\ast \mathrm{tr}_{A A^\prime}[( X \otimes Y ) (M_g {\otimes \mathds 1_B)}] V_g  \qquad \forall X \in \mathcal B(\mathcal H_A),~\forall Y \in \mathcal B(\mathcal H_{A^\prime B})
\end{equation}
and extended by linearity. 
Let us verify that indeed
\begin{equation}\label{teleportationprocessor}
\cP ( \rho^A \otimes c_T^{A^\prime B} ) = T(\rho).
\end{equation}
In the following, we insert the definition of the Choi-Jamio{\l}kowski state $c_T^{A^\prime B}=(\mathrm{id}_{A^\prime} \otimes T^{A^{\prime \prime}B}) \ket \Omega \bra \Omega_{A^\prime A^{\prime \prime} }$ and we use $(X \otimes \mathds 1) \ket \Omega = (\mathds 1 \otimes X^T) \ket \Omega$, $X^T = \bar X$ for Hermitian $X$ as well as $\tr_{A^\prime}[ (X \otimes \mathds 1) \ket \Omega \bra \Omega_{AA^\prime} (Y \otimes \mathds 1)] = 1/d_1 XY$. Here, $A^{\prime \prime}$ is again a system isomorphic to $A$. We calculate
\begin{equation}
\begin{split}
&\tr_{A A^\prime} [(\rho^A \otimes c_T^{A^\prime B} ) M_g \otimes \mathds 1_B]\\
&= d_1^{2}d\mu(g) \tr_{AA^\prime} \big[ (\rho^A \otimes ( \mathrm{id}_{A^\prime} \otimes T^{A^{\prime \prime}B}) \ket \Omega \bra \Omega_{A^\prime A^{\prime \prime} }) (\mathds{1}_{A} \otimes \bar U_g ) \ket \Omega \bra \Omega_{A A^\prime} (\mathds{1}_{A} \otimes \bar U_g )^\ast \otimes \mathds 1_B \big] \\
&= d_1^2 d\mu(g) \tr_{AA^\prime} \big[ \big( ( \mathrm{id}_{A^\prime} \otimes T^{A^{\prime \prime}B} )\ket \Omega \bra \Omega_{A^\prime A^{\prime \prime} } \big) (\mathds{1}_{A} \otimes \bar U_g ) (\mathds{1}_{A} \otimes \bar \rho^{A^\prime} )\ket \Omega \bra \Omega_{A A^\prime} (\mathds{1}_{A} \otimes \bar U_g)^\ast \otimes \mathds 1_B] \\
&= d_1^2 d\mu(g) \tr_{A^\prime} \big( ( \mathrm{id}_{A^\prime} \otimes T^{A^{\prime \prime}B}) \ket \Omega \bra \Omega_{A^\prime A^{\prime \prime} }  ( \bar U_g \bar \rho \bar U_g^\ast \otimes \mathds{1}_B)_{A^\prime} \big)\frac{1}{d_1} \\
&= \frac{ d_1}{d_1} T({U_g \rho U_g^\ast} ) d\mu(g) \\
&= V_g T(\rho) V_g^\ast d\mu(g),
\end{split}
\end{equation}
Applying the unitary conjugation as in Eq.~\eqref{def:teleportationprocessor} and integrating yields Eq.~\eqref{teleportationprocessor}. This shows that $\mathcal P$ is indeed a CPQP$_{UV}$ with the desired program dimension.
\end{proof}

Using the structure of the commutant $\mathcal K$, we can reduce the program requirements further:

\begin{theorem}\label{thm:teleportation-upper}
Let $U$ be an irrep {on $\cH_1$} of a compact group $G$, $V$ another representation {of $G$ on $\cH_2$}, and let $\mathcal K$ be of the form
\begin{equation}
\mathcal K = \bigoplus_{k=1}^K \mathds{1}_{b_k} \otimes \cB(\cH_k^\prime).
\end{equation}
Then, there exists a CPQP$_{UV}$ with program dimension $d_P = \sum_{k=1}^K n_k$, where $n_k = \dim(\cH_k^\prime )$.
\end{theorem}

\begin{proof}
Since we have a programmable quantum processor $\mathcal P$, constructed in Proposition \ref{prop:teleportation-processor}, available, we combine its protocol, i.e.\ the teleportation simulation part, 
with a compression map which reduces the program dimension. Instead of
$\mathcal K$ we consider a simpler matrix algebra with all multiplicities removed. These matrix algebras are isomorphic. Let
\begin{equation}
D^\ast : \mathcal K \to \bigoplus_{k=1}^K \cB (\cH_k^\prime)
\end{equation}
be this isomorphism and let $\cH_P = \bigoplus_{k=1}^K \cH_k^\prime$, which has dimension $d_P=\sum_{k=1}^K n_k$. Thus, $D^\ast$ is a unital completely positive map with unital completely positive inverse $C^\ast$~\cite[Example II.6.9.3(i)]{Blackadar2006}. Let $\tilde C$, $\tilde D$ be their dual maps. Both maps can be extended to quantum channels {$C: \cB(\cH_{1} \otimes \cH_2) \to \cB(\cH_{P})$, ${D}: \cB(\cH_{P}) \to \cB(\cH_1 \otimes \cH_2)$} by composing them with trace-preserving conditional expectations onto the respective subalgebras. Note that we identify the subalgebras in the images of $\tilde C$, $\tilde D$ with subalgebras of $\mathcal B(\mathcal H_{P})$ and $\mathcal B(\mathcal H_1 \otimes \mathcal H_2)$, respectively. The maps $C$ and $D$ are completely positive since the conditional expectations are~\cite[Proposition 5.2.2]{Benatti2009}. We define
\begin{equation*}
    \mathcal P^\prime = \mathcal P \circ (\mathrm{id} \otimes D),
\end{equation*}
and it follows that $\mathcal P^\prime(\rho \otimes {C}(c_T)) =T(\rho)$. 
\end{proof}

\begin{remark}
Comparing Proposition \ref{prop:teleportation-processor} to Theorem \ref{thm:teleportation-upper}, we see that the bound in the theorem is strictly better as soon as there is one index $k$ such that $b_k>1$. Recall that the $b_k$ are the dimensions of the irreps appearing in $\bar U \otimes V$. Thus, the improvement of Theorem \ref{thm:teleportation-upper} over Proposition \ref{prop:teleportation-processor} is larger, the more different irreps of dimension strictly larger than $1$ appear in $\bar U \otimes V$ and the larger the dimension of these irreps is.
\end{remark}

\section{Bounds for Approximate Programmability}
\label{sec:approximateprogrammability}
Since exact universal programmable quantum processors are impossible, there is a great interest in approximate versions of such devices. After having considered exact programmability in the last section, we now want to approximate the output such that the programmed output is close to the ideal output. Therefore, we need
 the concept of an $\epsilon$-CPQP$_{UV}$ {from Definition \ref{def:epsilonCPQP}}, which is a processor that implements a channel $T^\epsilon$ instead of the exact result $T$. However, it should be $\epsilon$-close to the ideal one {in diamond norm}. Having 
{recalled} this, the question of program requirements arises and hence, we show that the program {register} requirements for an $\epsilon$-CPQP$_{UV}$ are not much lower than for an exact CPQP$_{UV}$. 
Note that we can still benefit from the covariance property and the corresponding structure of the commutant ${\mathcal K}$.

\subsection{Upper bounds on the program dimension} \label{sec:approx-upper}
We construct generic upper bounds for the dimension of the program register {$d_P$}. We seek to establish an $\epsilon$-net on the set of covariant channels $\cT_{UV}$. Therefore, we need the following result about $\epsilon$-nets from Ref.~\cite{ledoux1991probability}.

\begin{lemma}[$\epsilon$-nets in $\mathbb R^n${~\cite[Lemma 9.5]{ledoux1991probability}}]\label{netlemma}
Let $\epsilon \in (0,1)$ and let $\|\cdot\|$ be any norm on $\mathbb R^n$. There is an $\epsilon$-net $\mathcal S$ on the unit sphere $S^{n-1}_{\| \cdot \|}$ of $(\mathbb R^n, \| \cdot \|)$ of cardinality
\begin{equation*}
    |\cS| \leq \left(1+\frac{2}{\epsilon}\right)^n.
\end{equation*}
That means, for all $x \in S^{n-1}_{\| \cdot \|}$, there is a $y \in \mathcal S$ such that $\| x-y\| \leq \epsilon$.
\end{lemma}

Due to the Choi-Jamio\l kowski isomorphism, there is a $c_T \in \cJ_{UV}$ corresponding to each $T\in \cT_{UV}$. 
According to Lemma~\ref{lemmaone}, we know that $\cJ_{UV} \subseteq \mathcal K \cap \cD(\cH_1 \otimes {\cH_2})$. Thus, we can make use of the special block-diagonal structure of $\mathcal K$. 

\begin{proposition} \label{prop:channel-net}
Let
\begin{equation}
\mathcal K = \bigoplus_{k=1}^K \mathds{1}_{b_k} \otimes \cB(\cH_k^\prime).
\end{equation}
For $\epsilon \in (0,1)$, there is a set $\mathcal S_{UV} \subseteq \mathcal T_{UV}$ such that for all $T \in \mathcal T_{UV}$, there is a $T^\epsilon \in \mathcal S_{UV}$ such that $\|T - T^\epsilon\|_\diamond \leq 2 \epsilon$. Moreover,
\begin{equation*}
    |\mathcal S_{UV}| \leq \left(1+\frac{2}{\epsilon}\right)^{d_n},
\end{equation*}
where $d_n = \sum_{k=1}^K n_k^2$.
\end{proposition}
\begin{proof}
Since $\mathcal J_{UV} \subseteq \mathcal K$ by Lemma~\ref{lemmaone}, the real vector space $\mathrm{Lin}_{\mathbb R}~ \mathcal J_{UV}$ generated by $\mathcal J_{UV}$ has dimension at most $d_n$. Using the Choi-Jamio{\l}kowski isomorphism, we infer that $\mathrm{Lin}_{\mathbb R}~\mathcal T_{UV}$ is a real subspace of $\mathcal B(\mathcal  B(\mathcal H_1), \mathcal B(\mathcal H_2))$ of dimension at most $d_n$. The restriction of the diamond norm turns $(\mathrm{Lin}_{\mathbb R}~\mathcal T_{UV}, \| \cdot \|_\diamond)$ {into} a real normed space which is isometrically isomorphic to $(\mathbb R^{d_n}, \|\cdot \|)$ for some induced norm $\|\cdot \|$. Moreover, $\|T\|_\diamond =1$ for all $T \in \mathcal T_{UV}$. Lemma \ref{netlemma} thus ensures the existence of an $\epsilon$-net $\mathcal S$ on the unit sphere of $(\mathrm{Lin}_{\mathbb R}~\mathcal T_{UV}, \| \cdot \|_\diamond)$ and the unit sphere contains $\mathcal T_{UV}$. To obtain $\mathcal S_{UV}$, we repeat the following steps. Take $\Phi \in \mathcal S$. If $\Phi \in \mathcal T_{UV}$, keep it and proceed to the next element. If there is no $T \in \mathcal T_{UV}$ such that $\|\Phi - T\|_{\diamond} \leq \epsilon$, remove $\Phi$ from the set and proceed to the next element. If there is a  $T \in \mathcal T_{UV}$ such that $\|\Phi - T\|_{\diamond} \leq \epsilon$, exchange $\Phi$ by $T$ and proceed to the next element. This algorithm constructs $\mathcal S_{UV}$ with the desired properties. Indeed, for any $T \in \mathcal T_{UV}$, there is a $\Phi \in \mathcal S$ and a $\mathcal T^\epsilon \in \mathcal S_{UV}$ such that $\|T - \Phi\|_{\diamond} \leq \epsilon$, $\|T^\epsilon - \Phi\|_{\diamond} \leq \epsilon$. Thus, $\|T^\epsilon - T\|_{\diamond} \leq 2\epsilon$. The cardinality bound follows since by construction $|\mathcal S_{UV}| \leq |\mathcal S|$.
\end{proof}
{The previous} proposition allows us to construct an $\epsilon$-CPQP$_{UV}$.
\begin{proposition}\mylabel{upperbound}
For a compact group $G$ and representations $U$ {on $\cH_1$}, $V$ {on $\cH_2$} such that 
\begin{equation}
\mathcal K =\bigoplus_{k=1}^K \mathds{1}_{b_k} \otimes {\cB(\cH_{k}^\prime)},
\end{equation}
there exists an $\epsilon$-CPQP$_{UV}$ with program dimension 
\begin{equation*}
    {d_P} \leq \left(1+\frac{2}{\epsilon}\right)^{d_n}
\end{equation*}
where $d_n = \sum_{k=1}^K n_k^2$ {and $n_k = \dim(\cH_{k}^\prime)$.}
\end{proposition}
\begin{proof}
Let $\mathcal S_{UV} = \{T^\epsilon_1, \ldots, T^\epsilon_s\}$ be the set from Proposition \ref{prop:channel-net}. Then, we can define a processor by
\begin{equation*}
    \mathcal P(X \otimes Y) = \sum_{i = 1}^s \langle i | Y | i \rangle T^\epsilon_i(X) \qquad \forall X \in \mathcal B(\mathcal H_1),~\forall Y \in \mathcal B(\mathcal H_P)
\end{equation*}
and extending by linearity. Here, $\{\ket{i}\}_{i \in \{1, \ldots, s\}}$ is an orthonormal basis {of $\cH_P$}. Choosing the program state for any $T \in \mathcal T_{UV}$ to be $\ket{i}\bra{i}$ if $\|T - T^\epsilon_i\|_\diamond${$\leq 2\epsilon$}, the map $\mathcal P$ can be checked to be an $\epsilon$-CPQP$_{UV}$ using Proposition \ref{prop:channel-net}.
\end{proof}

\begin{remark}
\label{rmk:nets-optimal}
Note that in the case in which $U$ and $V$ are the trivial representation and $d_1 = d_2 = d$, Proposition \ref{upperbound} states that
\begin{equation*}
  d_P \leq   \left(1+\frac{2}{\epsilon}\right)^{d^4}.
\end{equation*}
This does not match the upper bounds of $d_P \leq (K/\epsilon)^{d^2}$ obtained in Ref.~\cite{Kubicki19}, which are also derived using $\epsilon$-nets. The reason is that the construction in Ref.~\cite{Kubicki19} only implements all unitary channels in dimension $d$ instead of all quantum channels, which is what our construction does. In the same setting, a slight arithmetic improvement of the work~\cite{Renner20} yields lower bounds of the form
\begin{equation} \label{eq:yang-lower-bound}
    d_P \geq \left(1+ \frac{\Theta(d^{-2})}{\sqrt{\epsilon}}\right)^{{d^2-1}-\delta}
\end{equation}
if the program states are required to be pure, for any $\delta > 0$. Thus, the question remains if the bounds in  Proposition \ref{upperbound} are optimal for trivial symmetries. We will argue that this is not the case.

By Ref.~\cite{Choi75}, the extreme points of $\text{CPTP}(\cH_1,\cH_2)$ have Kraus rank of at most $d_1$, whereas general channels can have Kraus rank up to $d_1d_2$. The set of $\text{CPTP}(\cH_1,\cH_2)$ is a subset of a real vector space of dimension $d_1^2(d_2^2-1)$, thus by Carath{\'e}odory's theorem, any channel is a combination of at most $r=d_1^2(d_2^2-1)+1$ extremal channels (see also Remark \ref{rem:pure-mixed}). Thus, we only need an $\epsilon$-net on the set of channels with Kraus rank at most $d_1$. Let $N$ be the cardinality of this net. Then, we can approximately program all extremal channels with
a measure-and-prepare processor and pure orthogonal states in an $N$-dimensional program register. This implies that the same processor approximately programs all channels by using mixtures of up to $r$ basis states as program registers; alternatively, we could use the corresponding superpositions of the basis vectors to obtain pure program states. It remains to bound $N$.

We can construct the $\epsilon$-net on the set of channels with Kraus rank at most $d_1$ by considering their Choi-Jamio{\l}kowski states, which are states on $\cH_1\otimes\cH_2$ of rank $\leq d_1$. These can be purified to pure states on $\cH_1\otimes\cH_2\otimes\cH_1'$, where $\cH_1'\simeq \cH_1$. In Ref.~\cite[Lemma III.6]{aspects} it is shown that a $\delta$-net on such states with respect to the trace norm exists that has $N\leq\left(\frac{5}{\delta}\right)^{2d_1^2d_2}$ elements; tracing out $\cH_1'$ gives us the desired net on Choi-Jamio{\l}kowski states with bounded rank, since the trace norm satisfies data-processing. By Ref.~\cite[Eq.~(3.414)]{Watrous}, for two channels $T_1$, $T_2 \in \mathrm{CPTP}(\mathcal H_1, \mathcal H_2)$ and their respective Choi-Jamio{\l}kowski states $c_1$, $c_2$, it holds that
\begin{equation*}
    \| T_1 - T_2\|_\diamond \leq d_1 \| c_1 - c_2\|_1.
\end{equation*}
Thus, choosing $\delta = \frac{\epsilon}{d_1}$ provides the desired $\epsilon$-net.

We conclude that  Proposition \ref{upperbound} is not optimal, because it gives an exponent of $d^4$, whereas the above construction improves this to $O(d^3\log d)$. It remains open, however, if the latter is optimal or whether it is possible to further improve this exponent. Note that for mixed program states, the bound in Eq.~\eqref{eq:yang-lower-bound} does not apply, but we can purify the program states at only little extra cost, cf. Remark \ref{rem:pure-mixed}.
\end{remark}

\subsection{Lower Bounds for the program dimension}\label{sec:lowerbounds}
We seek to provide lower bounds {on} the program dimension of an $\epsilon$-CPQP$_{UV}$. The main idea is that all information about the $UV$-covariant channel $T \in \cT_{UV}$ is contained in its corresponding Choi-Jamio\l kowski state $c_T$. Thus, the program state $ \pi_T \in \mathcal D(\cH_P)$ has to store all information about $c_T$. Using the \textit{Holevo information} to quantify the amount of information, we obtain the following lower bounds:

\begin{theorem}\label{lowerbounds}
Let $\epsilon \in [0,1)$, $\cP^\epsilon \in {CPTP(\cH_1 \otimes \cH_P, \cH_2)}$ be an $\epsilon$-CPQP$_{UV}$, {$G$ be a compact group with an irrep $U$ on $\cH_1$ and $V$ be a representation on $\cH_2$}. 
Then, the following lower bound for the program dimension $d_P$ holds 
\begin{equation}
\frac{1}{2^{2 h(\epsilon)}} \bigg( {\sum_{k=1}^K} n_k \bigg)^{(1-2\epsilon)}
\end{equation}
with $h(\epsilon)= H(\epsilon, 1-\epsilon)=-\epsilon \log \epsilon - (1-\epsilon) \log (1-\epsilon)$ being the binary entropy, $n_k$ the multiplicity of the irrep $k \in \{1, \ldots, K\}$ in the direct sum decomposition of $\bar U \otimes V$
and $\epsilon$ the approximation parameter of the $\epsilon$-CPQP$_{UV}$.
\end{theorem}

\begin{proof}
Every channel $T^\epsilon$, the processor is able to implement,
corresponds to a Choi-Jamio\l kowski state $c_{T^\epsilon} \in \cD (\cH_1 \otimes \cH_{2})$ which can be understood as the output state of the processor $\cP^{\epsilon} \in$ CPTP$(\cH_1 \otimes \cH_P, \mathcal \cH_2)$ tensorized with the $d_1$-dimensional identity map $\mathrm{id}: \cB(\cH_1) \to \cB(\cH_1)$. The construction 
\begin{equation}
[\mathrm{id} \otimes \cP^\epsilon] (\ket \Omega \bra \Omega_{\mathcal H_{1} \otimes \mathcal H_{1}} \otimes \cdot) : \mathcal B(\mathcal H_P) \to {\mathcal B(\mathcal H_{1} \otimes \cH_2)}
\end{equation}
provides a quantum channel which maps every program state 
$\pi_T$ to a Choi-Jamio\l kowski state $c_{T^\epsilon}$.
This is a completely positive map because the processor map itself and the identity map are completely positive.

Let 
\begin{equation*}
\chi(\{\rho_i, p_i\}) := S\left(\sum_i p_i \rho_i\right) - \sum_i p_i S(\rho_i)    
\end{equation*}
be the Holevo information, where $S$ is the von Neumann entropy and $\{\rho_i, p_i\}$ is an ensemble of quantum states.
Let $p^{(k)}:=(p^{(k)}_1, \ldots, p^{(k)}_{N_k})$ be a probability distribution and let 
$\ket{\psi^{(k)}_j} \bra{\psi^{(k)}_j} \in \mathcal D_P(\mathcal H_{k}^\prime)$, $j \in \{1, \ldots, N_k\}$, be a collection of states such that $\{ \ket{\psi^{(k)}_j}, p^{(k)}_j\}$ is a $1$-design~\cite{Hayashi05, Ambainis07}, i.e.\ its average is the same as the uniform average over pure states with respect to the Haar measure. Then, {$p = (\lambda _1 p^{(1)}, \ldots, \lambda_K p^{(K)})$} is a probability distribution on Choi-Jamio\l kowski states $c_T \in  \mathcal J_{UV}$, where $p_j^{(k)}$ corresponds to $c_{T_{(k),j}} := 0 \oplus \frac{1}{b_k}\mathds{1}_{b_k} \otimes \ket{\psi_j^{(k)}} \bra{\psi_j^{(k)}} \oplus 0$. Here, $\lambda_k = \frac{n_k}{d_c}$, {where $d_c = \sum_{k=1}^K n_k$.} The $c_{T_{(k),j}}$ are constructed such that they are elements of $\mathcal J_{\mathcal UV}$ and their structure follows from the form of $\mathcal K$. It can be checked that the average state is 
\begin{equation*}
  \frac{1}{d_c}\overset{K}{\underset{k=1}{\bigoplus}} {\frac{\mathds 1_{b_k}}{b_k} \otimes \mathds 1_{n_k}.}
\end{equation*}
The probability distribution also induces ensembles {$\{c_{T^\epsilon_{(k),j}}, p_j^{(k)}\}$ and $\{\pi_{T_{(k),j}}, p_j^{(k)}\}$.}

Applying the data processing inequality~\cite[Section 10.7.2]{Wildebook} (since the Holevo information is {a} mutual information), we get
\begin{equation}
\chi(\{\pi_{T_{(k),j}}, p_j^{(k)}\}) \geq \chi(\{c_{T^\epsilon_{(k),j}}, p_j^{(k)}\}).
\end{equation}
Furthermore, note that the processor implements $T$ up to accuracy $\epsilon$, i.e.\ 
\begin{equation}
\frac{1}{2}\| T - T^\epsilon \|_\diamond \leq \epsilon
\end{equation}
which can be related to the corresponding Choi-Jamio\l kowski states~\cite[eq. 3.414]{Watrous}
\begin{equation}
\frac{1}{2}\| c_T  - c_T^\epsilon\|_1 \leq \frac{1}{2} \| T - T^\epsilon \|_{\diamond} \leq \epsilon.
\end{equation}
Note that $c_T$ has a block-diagonal structure inherited from 
\begin{equation}
\mathcal K =\bigoplus_{k=1}^K \mathds{1}_{b_k} \otimes \cB(\cH_{k}^\prime).
\end{equation}
Since $\bigoplus_{k=1}^K \mathcal \cB(\cH_k^\prime)$ and $\mathcal K$  are isomorphic as matrix algebras, extending the dual map of this isomorphism to a map $C:\mathcal \cB(\cH_{1} \otimes \cH_2) \to$ $\cB(\cH_{d_c})$ (as in the proof of Theorem \ref{thm:teleportation-upper}),
\begin{equation*}
     {C: \overset{K}{\underset{k=1}{\bigoplus}} \mathds 1_{b_k} \otimes B_k \mapsto  \overset{K}{\underset{k=1}{\bigoplus}} b_k B_k}
\end{equation*}
discards the multiplicity spaces and thus reduces the dimensions from $d_1 d_2$ to $d_c = {\sum_{k=1}^K} n_k$.
Since the trace distance is contractive under quantum channels,
\begin{equation*}
    \frac{1}{2}\|C(c_T)  - C(c_{T^\epsilon}) \|_1 \leq \frac{1}{2}\| c_T  - c_{T^\epsilon}\|_1\leq \epsilon.
\end{equation*}
With this trace-norm distance, we apply the Alicki-Fannes-Winter (AFW) inequality~\cite{Audenaert007, Petz},~\cite[Lemma 1]{Winter16} to bound the difference of the corresponding Holevo informations: 

\begin{equation}
\begin{split}
&\left| \chi(\{C(c_{T_{(k),j}}), p_j^{(k)}\}) -\chi(\{ C(c_{T^\epsilon_{(k),j}}), p_j^{(k)}\}) \right|\\ 
&= \left| S \left( \sum_{k j}C(c_{T_{(k),j}}) p_j^{(k)} \right) - \sum_{k j} S( C(c_{T_{(k),j}})) p_j^{(k)}  - S \left(\sum_{k j} C(c_{T^\epsilon_{(k),j}}) p_j^{(k)} \right)  +  \sum_{k j} S( C(c_{T^\epsilon_{(k),j}})) p_j^{(k)}\right|  \\
&\leq \left| S \left( \sum_{k j} C(c_{T_{(k),j}}) p_j^{(k)} \right) - S \left(\sum_{ k j} C(c_{T^\epsilon_{(k),j}}) p_j^{(k)} \right)\right| + \sum_{ k j} \bigl|S( C(c_{T_{(k),j}}))-S( C(c_{T^\epsilon_{(k),j}})) \bigr| p_j^{(k)}      \\
&\leq \epsilon \log {d_c} - h(\epsilon) + \epsilon \log {d_c} - h(\epsilon) \\
&= 2 \epsilon \log d_c - 2h(\epsilon),
\end{split}
\end{equation}
with $h(\epsilon) = H(\epsilon, 1-\epsilon) = - \epsilon \log \epsilon - (1-\epsilon) \log (1-\epsilon)$ the binary entropy, where we 
used the AFW inequality in the last step.  

Thus, we obtain
\begin{equation}
\begin{split}
\chi(\{\pi_{T_{(k),j}}, p_j^{(k)}\}) \geq &\chi(\{c_{T^\epsilon_{(k),j}}, p_j^{(k)}\})\\
\geq &\chi(\{ C(c_{T^\epsilon_{(k),j}}), p_j^{(k)}\})\\
\geq & \ \chi(\{ C(c_{T_{(k),j}}), p_j^{(k)}\})-2 \epsilon \log {d_c} - 2 h(\epsilon).\\
\end{split}
\end{equation}

Let us consider the term $\chi(\{ C(c_{T_{(k),j}}), p_j^{(k)}\})$. Since the $ C(c_{T_{(k),j}})$ of this ensemble are pure states, the second term of the Holevo information is zero. Moreover 
\begin{equation}
\sum_{k j} C(c_{T_{(k),j}}) p_j^{(k)} = \frac{\mathds 1_{d_c}}{d_c}
\end{equation}
and the von Neumann entropy of a maximally mixed state is $S \big( \frac{\mathds{1}_{d_c}}{d_c}  \big) = \log d_c$. Together with the inequality $\log d_P \geq 
\chi(\{{\pi_{T_{(k),j}}}, p_j^{(k)}\})$, we obtain
\begin{equation}
\begin{split}
\log d_P &\geq \chi(\{\pi_{T_{(k),j}}, p_j^{(k)}\})\\
& \geq (1- 2 \epsilon) \log d_c -2 h(\epsilon) \\
&= \log d_c^{(1-2\epsilon)} -2 h(\epsilon).
\end{split}
\end{equation}
This yields
\begin{equation}
d_P \geq \frac{1}{2^{2 h(\epsilon)}} d_c^{(1-2 \epsilon)}= \frac{1}{2^{2h(\epsilon)}} \bigg( \sum_{k=1}^K n_k \bigg)^{(1-2 \epsilon)},
\end{equation}
{which} proves the assertion.
\end{proof}

The bound on $d_P$ we have derived increases with the block dimension $n_k$.

\section{Discussion}\label{discussion}

 In this section, we discuss our findings and mention some potential future work. We start by pointing out that Theorem \ref{thm:teleportation-upper} is in fact optimal. Note that for $\epsilon=0$ we obtain exact lower bounds from Theorem \ref{lowerbounds}. Combining them with 
 the teleportation protocol in Theorem \ref{thm:teleportation-upper}, the optimal program requirements $d_P$ for an CPQP$_{UV}$ are thus
 \begin{equation*}
     d_c=\sum_{k=1}^K n_k = d_P.
 \end{equation*}
 If we insisted on pure program states, we would get $d_P = d_c^2$ from Theorem \ref{thm:teleportation-upper}, replacing $\pi_T$ by a suitable purification (for example the canonical purification $\sqrt{d_P}(\mathds 1 \otimes \pi_T^{1/2} W) \ket{\Omega}$, where $W$ is a suitable unitary). More refined constructions have been discussed in Remark \ref{rem:pure-mixed}. We leave it as an open question whether the upper bounds can be improved to match the lower ones also for pure program states. In the case where the commutant $\mathcal K$ is abelian, Corollary \ref{cor:abelian} answers the question in the affirmative.
 
 In the present paper, we have shown that the problem {of finding} programmable quantum processors for $UV$-covariant channels for irreducible $U$ is very different from the situation concerning universal programmable quantum processors. While the former is always possible exactly with finite-dimensional program register, the latter is only possible approximately with finite-dimensional program register and requires a large program dimension. The question remains whether there are intermediate situations:
     Is there a group $G$ and a reducible representation $U$ such that there exists an exact CPQP$_{UV}$ with finite-dimensional program register? If so, what determines if there are exact CPQP$_{UV}$ with finite-dimensional program register? A major roadblock to extend our results to the case in which $U$ is no longer irreducible is Lemma \ref{lemmatwo}, since we can no longer guarantee that all elements in $\mathcal K \cap \mathcal D(\mathcal H_1 \otimes \mathcal H_2)$ correspond to trace-preserving maps.

 A further question is, how upper bounds for $\epsilon >0$ can be constructed because {the} $\epsilon$-nets used in Ref.~\cite{Kubicki19} and the general-purpose result Proposition \ref{upperbound}, for instance, rely on orthogonal program states. Remark \ref{rmk:nets-optimal} already shows that the proposition is suboptimal for trivial symmetries. If $U$ is irreducible it can be outperformed applying the teleportation protocol as we saw in Section~\ref{sec:teleportation}. Hence, we have good exact upper bounds and for an approximation in the case where the program states are not orthogonal, a different way to compress the program states is required. Therefore, the question arises whether we can find approximate upper bounds which improve the exact ones from Theorem \ref{thm:teleportation-upper} for $U$ irrep. However, Theorem \ref{lowerbounds} shows us that there is not much space for improvement using approximate processors.

\myacknowledgements
AB and MG thank Matthias Christandl, Frank Himstedt, Alexander Müller-Hermes, Ion Nechita, Jitendra Prakash, Cambyse Rouzé, Daniel Stilck-Fran{\c c}a, Simone Warzel and Michael Wolf for discussions.
AB acknowledges support from the VILLUM FONDEN via the QMATH Centre of Excellence (Grant no.\ 10059) and from the QuantERA ERA-NET Cofund in Quantum Technologies implemented within the European Union’s Horizon 2020 Programme (QuantAlgo project) via the Innovation Fund Denmark. MG is funded by the Deutsche Forschungsgemeinschaft (DFG, German Research Foundation) under Germany’s Excellence Strategy – EXC-2111 – 390814868. AW acknowledges financial support by the Spanish MINECO (projects FIS2016-86681-P and PID2019-107609GB-I00) with the support of FEDER funds, and the Generalitat de Catalunya (project CIRIT 2017-SGR-1127).

\bibliographystyle{plainnat}
\bibliography{mybib.bib}
\end{document}